# "What do you expect? You're part of the internet": Analyzing Celebrities' Experiences as Usees of Deepfake Technology

Experiences of deepfake abuse imagery and recourses

John Twomey [a], Sarah Foley [a], Sarah Robinson [a], Michael Quayle [b] [c], Matthew Peter Aylett [d] [e], Conor Linehan [a], Gillian Murphy [a]

[a] School of Applied Psychology, University College Cork, Cork, Ireland

[b] Centre for Social Issues Research and Department of Psychology, University of Limerick, Limerick, Ireland

[c] Department of Psychology, School of Applied Human Sciences, University of KwaZulu-Natal, Pietermaritzburg, South Africa

[d] Mathematics and Computer Science, Heriot Watt University, Edinburgh, United Kingdom

[e] CereProc Ltd., Edinburgh, United Kingdom

Deepfake technology is often used to create non-consensual synthetic intimate imagery (NSII), mainly of celebrity women. Through Critical Discursive Psychological analysis we ask; i) how celebrities construct being targeted by deepfakes and ii) how they navigate infrastructural and social obstacles when seeking recourse. In this paper, we adopt Baumer's concept of "Usees" (stakeholders who are non-consenting, unaware and directly targeted by technology), to understand public statements made by eight celebrity women and one non-binary individual targeted with NSII. Celebrities describe harms of being non-consensually targeted by deepfakes and the distress of becoming aware of these videos. They describe various infrastructural/social factors (e.g. blaming/silencing narratives and the industry behind deepfake abuse) which hinder activism and recourse. This work has implications in recognizing the roles of various stakeholders in the infrastructures underlying deepfake abuse and the potential of human-computer interaction to improve existing recourses for NSII. We also contribute to understanding how false beliefs online facilitate deepfake abuse. Future work should involve interventions which challenge the values and false beliefs which motivate NSII creation/dissemination.

## 1 INTRODUCTION

Deepfake technology is the latest example in a long history of technology facilitating violence towards women [29]. Deepfakes involve the use of deep learning technology to create false images/videos of their targets. The vast majority of deepfake videos available online are pornographic [1], and the people appearing in those videos often have not provided any form of consent. Deepfake technology is underpinned by infrastructure that is implicitly tied to the abuse of celebrity women [88], including software repositories and forums that produce non-consensual deepfakes of celebrities. Even supposedly positive uses of the technology in entertainment (such as the consensual resurrecting and de-aging of actors) often use software such as DeepFaceLab, which directly advertises forums dedicated to the production/dissemination of illegal content [77].

Most existing HCI research on deepfakes focuses on experimentally investigating perceptual differences in people's ability to identify deepfakes [54, 70, 89]. There has also been an increasing focus on understanding the social context behind deepfakes and implications on society. For example, a synthesis of qualitative studies on deepfakes by Vasist and Krishnan has criticized existing literature for studying deepfakes in isolation without considering how deepfakes are encouraged by platforms [82]. Their work also highlighted how much existing qualitative research on deepfake abuse imagery studies the platforms and communities which create the imagery as opposed to the experiences of victims. The limited qualitative work on deepfake abuse imagery from a human-computer interaction (HCI) perspective similarly focuses on understanding general public attitudes towards deepfake abuse material online, often through studying platforms such as Reddit [24] and asking the broader general public on their experiences with creating and viewing deepfake abuse imagery [81]. Outside of HCI, other qualitative approaches to this topic highlight the lived experiences of deepfake abuse victim-survivors [68] but do not situate it in the broader context of novel technologies and how they interact with online infrastructure and communities. Our work focuses specifically on qualitatively analyzing the statements of victims of deepfake nonconsensual synthetic intimate imagery (NSII) as a way of understanding current recourses and critical issues with AI technologies.

In the current paper, we examine the infrastructure of existing recourses available, and identify that responsibility is unfairly placed on the victims of deepfakes to remove abusive content. Through qualitative methods we highlight that the discourse employed by deepfake victim-survivors provides insights into their experience of manipulated intimate imagery. We analyze how victims identify infrastructural factors in the production, distribution, and consumption of deepfake abuse imagery (or non-consensual synthetic intimate imagery- NSII). We argue that the celebrity victims of deepfakes should be viewed as Usees of the technology as defined by Baumer [7]: stakeholders who are non-consenting, unaware and directly targeted by technology. Celebrities are intentionally targeted by the deepfake software, they are often initially unaware that there are deepfakes made of them, and the videos/generative models override their consent. It is important to understand how these high-profile figures have experienced being Usees of deepfake technology as they are most often the targets of deepfake abuse. Examining these public statements may also more broadly help the design of general solutions/interventions for deepfake abuse which will become increasingly relevant as more people are targeted.

This qualitative study will analyse public interviews with celebrity targets of deepfake intimate imagery. Our research questions are:

- How do celebrity victims of deepfake abuse construct being Usees of deepfake technology (the experience of being unaware and non-consenting targets of technology)?
- How do they negotiate infrastructural and social obstacles against seeking recourse for deepfake abuse?

We analyzed public interviews with celebrity targets of deepfakes using a form of Critical Discourse Analysis [20] known as Critical Discursive Psychology which is concerned with both the 'micro' analyses of discourse construction of discursive psychology alongside 'macro' considerations of societal injustice associated with Critical Discourse Analysis [41]. Our research questions are inherently discursive as they concern the positioning and construction of deepfake targets and their experiences with deepfake content, as well as how current solutions are framed (such as removing content from search engines/social media, and potential legal actions against the creators/consumers of deepfakes). By taking these interviews as a corpus of public statements of the harms of deepfake abuse, we highlight the real-world experiences of current targets. Their status as celebrity women and non-binary individuals means that they 1) are uniquely targeted and victimized by deepfakes as a pattern of abuse against public women and gender minorities, and 2) provide insights into harms of technology which may soon become widespread, as deepfake technology becomes increasingly accessible. The contribution of our work is to critically examine how technology infrastructure and usage patterns encourage deepfake abuse, how this is understood by the Usees of deepfake technology in their interaction with this infrastructure, and how this can inform software engineering, policy and interventions.

### 1.1 Note on terminology used

As HCI researchers it is important to be aware of the evolving terms around gendered violence to avoid stigmatization of survivors. There is limited guidance on the correct terminology to use for deepfake videos using manipulated intimate imagery in existing literature, despite criticisms of the phrase "deepfake pornography" [68], stemming from similar criticism over the phrase "revenge pornography" [48]. Existing terminology tends to imply sexual gratification and ignores both the harm and lack of consent. This paper refers to deepfake videos made from pornographic videos as an example of non-consensual synthetic intimate imagery (NSII) [81], or as deepfake abuse, drawing from the current terminology for image-based sexual abuse [58]. Similarly, there is limited guidance on the preferred terminology for those who have experienced deepfake abuse. When drawing from Baumer's work [7], we use the terms Usee or target. Otherwise, we use victim, while understanding that as this harm evolves, so will the preferred terminology.

## 2 RELATED WORK

Deepfake technology generally involves audiovisual media that has been manipulated using AI to falsely portray a person in a situation that has not occurred [84]. While deepfakes aren't the first method for manipulating video, they are significantly more accessible than manually editing video footage. From their inception, deepfake videos have been used in the targeted harassment of public women. Deepfakes originated in late 2017, evolving from a previous internet practice of photoshopping celebrities' likeness onto adult images (on the subreddit r/celebfakes and on sites such as 4chan) [11, 12, 74]. A Reddit user under the pseudonym u/deepfake, applied novel video manipulation technology to create similarly modified adult videos which he shared on a new subreddit [12]. While the resulting subreddit for these videos (also bearing the name deepfakes) introduced this technology to a wider audience, it should be noted that similar photoshopped NSII (which were not generated using AI) existed on Reddit and other parts of the internet for years beforehand. The community was characterized as both a space for learning, where its users gave insight and technological advice on the production of the videos [24], and as a place for the dissemination of NSII, specifically modified pornographic videos into which a celebrity's face was interpolated [77]. That forum can be seen as an example of an online environment characterized by "networked misogyny" which combines the anonymity of most internet forums with organized harassment and hostility against women [45]. An investigative news piece in January 2018 increased public awareness of deepfakes and led to the original deepfake subreddit and r/celebfakes being shut down [30]. Since then, deepfake creators have migrated to websites specifically designed for disseminating and creating manipulated NSII. These sites operate similarly to the original subreddit: with sub-forums categorized into general discussion, technical resources and assistance, creation resource sharing, content sharing, content requests and meta discussion [77]. These forums are part of the increasing number of deepfake marketplaces online, who produce deepfakes of famous and non-famous targets for a paid fee [68, 77].

## 2.1 Celebrity Women as Deepfake Usees

Research has shown that celebrity victims of NSII are viewed less sympathetically than non-celebrities, especially when the celebrity is a woman [22]. This is consistent with viewing behaviors of deepfakes: research by Umbach and colleagues [81] who found that 9.9% of men and 2.8% of women reported viewing sexual deepfakes of a celebrity (versus 4.2% and 1.6% respectively for deepfakes of ordinary people). Feminist literature suggests that celebrities exist in a fragile and tenuous space between the private and the public where their career involves constant scrutiny on their identity and actions [58]. Celebrities (particularly celebrity women) have always faced constant surveillance and significant breaches to their privacy and personal agency [28]. They are assumed to have traded control over their image to the public in return for their continued existence as cultural figures [59]. Information which is not public is considered a secret to be traded and sold as gossip [59]; this practice has only evolved with the advent of social media, with notorious publishing of private images by black hat hackers [55]. Feminist scholars have portrayed this violation of privacy as part of a distinct form of misogynistic harassment [58]: "Her images are stolen and circulated with a dual purpose: to both punish her for her public success, and to remind all women that their bodies/sexuality are never theirs, and are always prone to such violence, no matter how "successful" they become". Deepfakes build on this in a literal sense, where the body and image involved does not belong to the target at all but is instead the result of a fabrication.

While deepfakes are in theory able to target anyone, most victims remain public figures, and are almost exclusively women [37, 77]. A report by the cybersecurity company Deeptrace found in 2019 that 95% of deepfakes online were manipulated pornographic content and that the prevalence of deepfake content had doubled since the previous year [1]. This is a growing and global phenomenon. In recent years there has been an overwhelming increase in the targeting of K-pop idols, who now constitute around 25 percent of deepfake NSII online [83]. The targets have also broadened to various forms of online celebrity including social media influencers and streamers (users of live-streaming/vlogging platforms such as Twitch) [34]. As this technology becomes increasingly accessible, its potential to harm also broadens outside of celebrity targets. In Autumn 2024, the messaging company Telegram was investigated by South Korean police for hosting a group dedicated to deepfake NSII with over 200,000 members [47]. This was part of a larger investigation into deepfake groups within universities, where students created NSII targeting fellow students and teachers [72].

In this work, we configure the targets of deepfakes within Eric Baumer's work on Usees [7] and post-userism [8]. Previously, HCI research typically conceived of individuals as belonging to one of two groups, the users and non-users of a given technology. HCI as a discipline was seen as concerning itself with users of technology, distinguishing the field from related disciplines. Baumer's work emerged from an increasing focus by HCI researchers on "non-use" of technology and contributes the novel idea of "Usees" of technology; a form of interaction which is neither use nor non-use *of*, but rather use *by*, technology. Usees are non-consenting and unaware persons who have been targeted by a technology or its users, a relationship with technology which is becoming more common. One example of Usees given by Baumer is that of stalkerware applications, where Foursquare's API was being used to track and map women without their consent [7]. We suggest that the targets of deepfake videos can also be described as Usees, as the content is made without their consent or awareness, and the software is designed to target a specific group: public women. Our research aims to explore how non-consent, unawareness and being targeted underlie the experience of deepfake abuse.

## 2.2 The Consequences of Being a Deepfake Usee

We situate this work within existing HCI literature which aims to explore the role of technology and social media in encouraging gendered abuse, such as work analyzing the sentiment and hatefulness of the #Gamergate movement, which involved the harassment and defamation of female video game journalists and developers on Twitter [13]. HCI work has also identified the importance of social media in challenging gendered abuse through activism online. Research focusing on the #MeToo movement showed how #MeToo encouraged Twitter users to talk about their experience with sexual violence and that the network of disclosure reduces stigma and encouraged other twitter users to speak out [23]. Yet divergent opinions on this exist which highlight the voices in these movements who were either unable to speak out or who viewed activism as a last resort for justice and were disappointed if this wasn't successful. This was a significant contribution of interviews with Indian activists and stakeholders carried out by Moitra and colleagues, who describe the marginalization of intersectional identities in activism and how the #MeToo movement served as a point of contact between legal and digital justice infrastructures [52]. There is limited work exploring the experiences of victims of deepfake abuse and the various socio-cultural and infrastructural obstacles which they face. Noelle Martin, a deepfake victim-survivor turned activist/scholar, describes two distinct phases of deepfake abuse: the "abuse phase" and the "removal phase" of deepfakes [43]. The lay advice given to victims of image-based sexual abuse (IBSA) often puts responsibility onto the victims; suggesting that they should reduce their public profile online and keep images of themselves off the internet [36, 42]. We refer to this kind of public advice as "social solutions" in this text as solutions which require victims of abuse to change their own behaviors online. The "strategy" of removing oneself from the internet to prevent deepfake abuse is a form of victim blaming which deprives women (and particularly already marginalized groups such as women of color) equal access to the internet [43]. Social solutions are also uniquely impractical for celebrities, (particularly internet celebrities) because they further threaten the labor undertaken by women and marginalized groups in the digital creative economy, which is already precarious under platform capitalism [16]. This strategy also emboldens the use of deepfakes in targeted harassment, where certain misogynistic communities online can police public women's behavior and force them to withdraw from the internet [57] (such as Gamergate [36, 58]).

Just as social solutions put the onus of the work on the victims, technological recourses for deepfakes often require users to self-police illicit content of themselves [81]. In particular, Reddit practices towards harmful content have been criticized because, while problematic communities can be banned, individual infractions overly rely on user-level interventions, such as community moderators and user reporting [25]. Existing policies from search engines such as Google Search [65] and Microsoft Bing [50] put the onus on users (or a representative) to collect screenshots of abuse imagery of themselves in their reports. This process must then be repeated for every search engine provider and does not take any action against the potential for future infractions (which may be increasingly common as the technical skills necessary to create deepfakes have reduced [40]. While this process is an improvement over the lack of any reporting mechanisms, the emphasis on user-based solutions is part of a wider issue with responsibility on platforms, where certain websites/services which are responsible for deepfakes ignore their own culpability and face limited consequences for hosting illicit content. Instead, they pass responsibility onto the victim to remove this content. The design of these services is a space in which HCI and particularly trauma informed HCI practice is required.

There are also legal recourses for deepfake abuse imagery though, like most technological solutions, they are also limited by focusing victims reporting known abuse and abusers. Intimate imagery generated through deepfake technology, like other forms of online non-consensual intimate images is a technology-facilitated crime [21]. There are various legal approaches that can be taken such as copyright and IP law [80], privacy [27] and defamation [57] and image rights [17]. As with all image-based sexual abuse, there are various obstacles to victims achieving justice [86]. Unfortunately, most legal solutions to deepfake abuse imagery are further hamstrung by two specific factors. Firstly, the anonymity of the creator impedes prosecution. Secondly, the transnational nature of internet crime [32] means that even if perpetrators were identified, they may be operating in an area of the world with different or non-existent deepfake regulations. Once the imagery exists it can be copied, shared and stored across various locations internationally [76]. Existing research on deepfake abuse forums shows that many deepfake perpetrators know this, and while they are concerned at the possibility of being prosecuted for producing illegal content, they believe that the content will always exist as long as they have access to the open-source technology behind the videos [77].

As part of the potential of HCI research in challenging gendered violence, one must center the experiences of those who have experienced abuse and how their abuse is impacted by technology [39], such as the development of care infrastructures for victims of domestic abuse, to respond to the threats they face to their privacy and security [78]. There is limited qualitative work on deepfake abuse which centers the victims of NSII. An interpretative phenomenological analysis (IPA) carried out by Rousay [68] on both online interviews with deepfake abuse victims and published work of lived experience experts/activists generated three common themes. Firstly, victims reported experiencing gendered narratives around deepfake victimization, feeling denigrated and objectified, and experiencing gendered stigma when reporting the abuse. Secondly, victims reported long-lasting consequences including post-traumatic distress, anxiety and nightmares. They also experienced harassment including doxing, rape threats and 'slut-shaming'. Finally, victims noted an effect on their identity where their reputation had changed and they felt like they had "become pornography". Existing qualitative research outside of HCI focuses on highlighting the experience of victims of deepfake abuse. Building on this, our work seeks to use Critical Discursive Psychology to explore more critically how these experiences are constructed in relation to broader social repertoires and how they emerge as a consequence of the infrastructure of deepfake technology, recourses for NSII, and online culture as a whole. Specifically, our work seeks to use a Critical Discursive approach to 1) examine the discourse around how technology infrastructure and usage patterns fuel abuse, and 2) how victim-survivors understand, respond to, interact with, and critique that infrastructural environment when seeking recourse.

## 2.3 The Infrastructural Justice Approach to NSII

In the last twenty years, digital communication technologies have been embedded into varying existing and novel forms of gendered violence and abuse [61]. This includes the abuse of public women using non-consensual intimate imagery. Feminist scholarship on the early 2000s- mid 2010s culture of sharing illicit and leaked images of female celebrities have highlighted how misogynistic targeting of celebrity women is constantly evolving: "the distaste for the public woman is not new, but what is new is the capacity for and techniques used to humiliate and punish her (p.45)" [58]. Recent work by Robinson and colleagues [67] suggest that deepfake technology is a form of infrastructural injustice, where the harms of this technology are created by many people and things operating within established norms and practices. Robinson et al [67] frame this in line with Susan Leigh Star's infrastructure, which only becomes infrastructure 'through use'; in the case of deepfakes, through the development of deepfake software and the hosting and distribution of deepfake content.

A focus on infrastructure moves focus from the Usee, back to the developers of deepfake content and perpetrators of deepfake abuse, and the technological infrastructure that enables abuse. Robinson and colleagues focus specifically on four aspects of Susan Leigh Star's infrastructure as applied to deepfake abuse imagery: learned as part of membership, links with conventions of practice, embodies existing standards and is built on an installed base (fig.1).

Figure 1: Summary of Robinson's [67] work on the infrastructure behind the creation of deepfake NSII

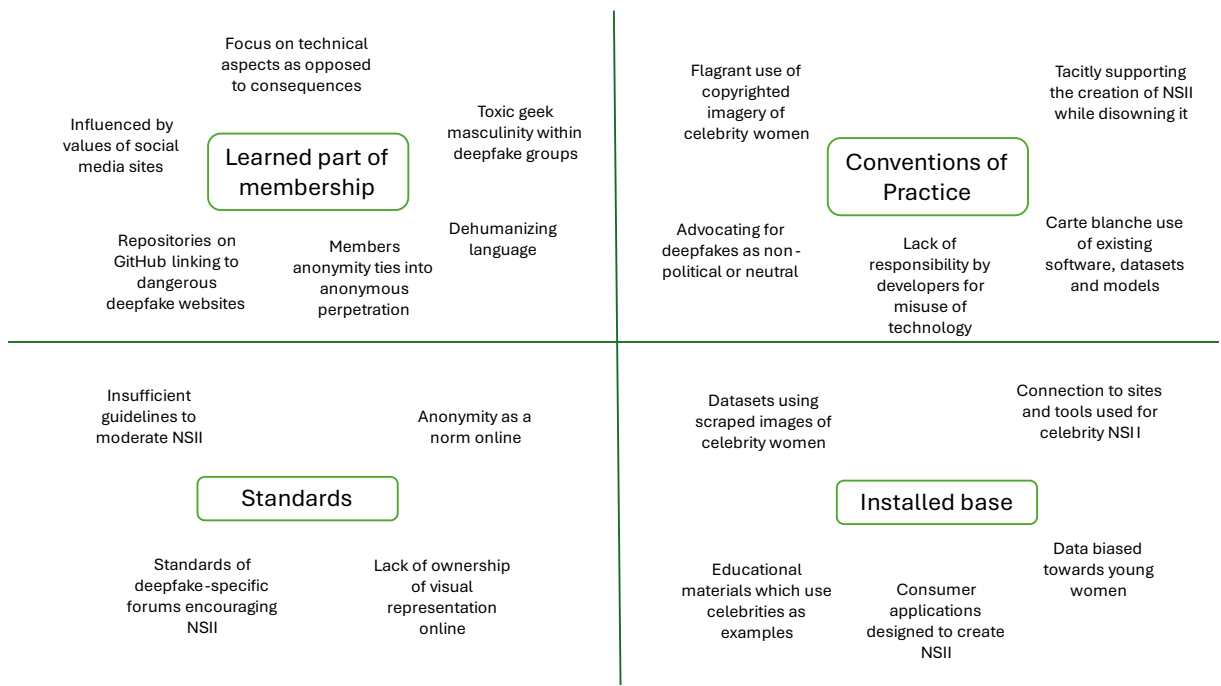

Firstly, deepfake software emerges from a community of practice (forums) and neutral/misogynistic attitudes towards deepfake targets/content are learned as part of membership of that community. Secondly, these communities both perpetuate and create conventions of practice: the ethos of open-source software views the software itself as neutral and encourages making it accessible to as many people as possible. Thirdly, it embodies the standards of internet services towards popular but immoral communities, particularly those of illegal content such as piracy, which operate under systems of moderation which are not designed to remove the illegal content until forced to. Finally, and perhaps most pertinently, is the installed base behind deepfakes (the pre-existing structure/s such as trained models, software and datasets). Today some of the most popular open source deepfake technology is linked overtly to websites specializing in the production of NSII targeting celebrity women [88], and still come with pre-trained celebrity models (often of young women) [46]. These models come with a disclaimer that they do not resemble any real person, but they clearly resemble celebrities with slightly modified pseudonyms. For example, if your name was Jamie McDonald, it would have a picture of what is clearly your face with the name Amy Mac Ronald underneath. Consumer deepfake apps try to have it both ways, e.g. they outlaw pornographic content but do allow users to swap faces onto the cover page of Playboy magazine [63]. The developers absolve themselves of responsibility for the content while they tacitly encourage the creation of non-consensual intimate imagery of women.

Though existing literature has charted various avenues for support for the victims of deepfakes such as self-reporting abuse materials, it is apparent that most solutions requiring self-report place part (or all) of the responsibility for reporting and removing content onto the target themselves. A distinct part of being a Usee of a technology is the lack of awareness, yet much of the existing recourses rely on victims being aware that this content exists before having to report it. Robinson et al.'s work [67] on deepfakes and infrastructural justice highlights how power, privilege, interest, and ability affect the development of solutions to the fundamental issues of deepfake abuse. These same principles apply to existing solutions; legal approaches have an interest in reducing the harms of deepfake abuse but lack the power and ability to do so. Conversely, search engines and internet service providers may have the power and ability to remove content but do not take more proactive solutions (presumably due to lack of interest as a whole, though individuals within these organizations may be actively working to improve recourses). They privilege the abuser over the abusee by providing the former with a platform/anonymity and the latter with the experience of having to diligently collect evidence of their own abuse (by emphasizing manual self-report as the main solution offered to victims of NSII). Social solutions also privilege the abuser by encouraging victims to remove themselves online which just facilitates the abuse (by showing that it is a successful strategy to force activists and public figures to self-censor). We do not propose further novel solutions here, but we seek to demonstrate why current recourses are inherently infrastructurally stymied by the values of Usee-centered technology. Studying the experiences of the current victims of deepfake abuse with existing recourse helps to highlight their limited efficacy and exposes their role in contributing to the overall distressing effects of deepfake NSII.

## 3 METHODS

### 3.1 Data collection

To carry out research on the experiences of celebrities with deepfake intimate imagery we took public statements and news interviews as case studies. As celebrities are most commonly the target of deepfake abuse materials we wanted to examine their experiences [77]. As well as this, using public statements as case studies is an appropriate methodology for HCI research on distressing topics as it avoids risks of re-traumatizing interviewees [49]. Furthermore, existing Critical Discourse Analysis literature has suggested the benefits of using a wide variety of existing sources for a corpus, especially from new forms of media communication such as social media and various forms of online content creation [38]. We identified numerous celebrities who had been mentioned as victims of deepfake imagery using a dataset from our previous scoping review studying extracts from the academic literature on harms of deepfakes [79]. In April 2024, we carried out a google keyword search using variations of the phrase "{name} speaks out deepfakes" to locate public statements. While the terminology of our search was specifically chosen to avoid abuse websites, pornographic websites occasionally still emerged as suggested search results – further highlighting the extent to which deepfake media is platformed by our information environment. Between April and June 2024, an additional search of online news articles about deepfake NSII was also conducted to identify further victims (i.e. searching Google and X using the phrase *speaks out deepfakes* or *celebrity speaks out deepfakes* to find relevant news articles and quotes). Interviews within larger texts were excluded if they were too short (e.g. a single sentence in a larger text) or heavily editorialized (e.g. where a newspaper article has worked direct quotations directly into the text), to avoid the risk of taking quotes out of context. As expected from qualitative work on pre-existing public statements, the texts collected were a wide range of lengths and sources. Outside of issues with brevity and editorialization, we wanted to include as large a number of celebrities as possible in this dataset as is best practice to avoid criticisms of arbitrarily selecting text [73]. The majority of statements were between 200 -1000 words long. We included multiple statements from celebrities in cases which they had made multiple statements. Further details on the corpus are presented below in table 1.

We analyzed both text and video responses. Audio was transcribed using the transcription tool in Microsoft Word, which was manually refined by the first author. The total size of this corpus was 19,202 words, from 9 celebrities across 12 sources (Table 1) which is an appropriate sample size as discourse analysis methods acknowledge that large sizes of sample and text would make the analytic task unmanageable as opposed to improving the quality of an analysis [10]. Regarding demographics, while the chosen interviews reflect that the majority of early targets of deepfakes were celebrity women from English speaking countries [77] we did attempt to find instances of other groups of celebrities (such as Bollywood actors and K-pop idols) who had spoken out about deepfake abuse imagery but could not find any such public statements at the time of data collection. One of the celebrities in the dataset is non-binary and has expressed that they do not have a specific preferred pronoun (E). There are inherently issues when studying secondary data and attributing demographic information that participants have not consented to self-disclose. Though we are not able to give demographic data we did consciously seek to represent celebrities that come from a broad range of ethnicities, gender identities, social classes and sexuality. While not appropriate for our study of secondary data, we recognize the need for work which situates experiences with deepfake NSII directly within intersecting social and cultural factors and that further considering race, gender and sexuality would offer a complementary lens for such an analysis.

Ethical approval was obtained from our institution's ethics board. This topic requires sensitivity to the experiences of the individuals involved. While the original data is publicly accessible online, we have chosen to obscure the celebrities in our corpus. The issue of anonymity/identification split the research team: originally, we intended on including the identity of the public interviewees because we felt that removing their names would be against the spirit of speaking out. However, when analyzing the data we found that many expressed concerns about their names/brand being associated with deepfakes in perpetuity.

Table 1: Deepfake statements in our corpus

| Title | Type of celebrity | Description and word count |
|---|---|---|
| A | Streamer, activist | Tweet (203) |
| B | Streamer | One Vlog (394), one interview (1029) and one podcast (6831) |
| C | Streamer | Video (1410) and tweet (51) about experience of deepfakes |
| D | Influencer | One TikTok post (412) |
| E | Streamer | Video about experience with deepfakes (691) |
| F | Actress | Text statement for news article (425) |
| G | Actress | Two extracts from same interview (318) |
| H | Actress, influencer | One talk show interview (348) |
| I | Rapper | Tweet (39) |

**3.2 Analysis**

This work uses Critical Discursive Psychology as its method of qualitative analysis. Critical Discursive Psychology draws on both discursive psychology, which is interested in discursive practices i.e. what people do with language, and poststructuralist Foucauldian discourse analysis which is interested in "what kind of objects and subjects are constructed through discourses and what kinds of ways-of-being these objects and subjects make available to people." [87]. Potter and Wetherell [60] suggest that combining these approaches is preferable. "As CDA focuses on discourses, which position people along power lines, CDP focuses on interpretative repertoires which 'allow for a wider focus on human agency' and are 'recognizable way[s] of describing, framing or speaking about an issue that is identifiable as such (p.240)" [41]. Outside of interpretative repertoires we also study ideological dilemmas where people construct a struggle with divergent constructions of social relationships and realities [20]. By further engaging with Critical Discursive Psychology literature [31], we were able to consider how identity, memory and trauma were framed through the discourse of the corpus.

We chose Critical Discursive Psychology as it draws from Critical Discourse Analysis [20] to study the role of discourse in broader social wrongs/injustices – such as those which underly technology facilitated gendered violence towards female celebrities. This allows researchers to take a critical realist approach [71] in how the constructed discourse links to the larger material concerns of an issue. For us, this meant analyzing how discursive techniques give insight into the experiences of deepfake such as framing and positioning (accepting or rejecting one of many subject-positions which fit into broader interpretative repertoires [85]). While outside the analytic lens of discourse analysis, we did consider narratives in the text in how the narrative of their abuse/harassment is constructed in terms of larger narratives of technology fueled gendered violence. We incorporated Baumer's concept of Usees into the analytic lens by specifically considering how non-consent was constructed and framed within interpretative repertoires of other non-consensual acts, how celebrities framed the experience of being targeted through broader lenses of gendered violence, and finally how they constructed awareness and the distress of becoming aware of NSII.

Before carrying out the analysis, it was necessary to familiarize ourselves with current legislation for deepfake intimate imagery. The first author compiled a document outlining the laws around deepfakes in the countries where potential celebrities lived, and we drew from comparative literature outlining the variety of current legal, technical and social solutions. This helped contextualize the experiences in the corpus within existing legal and technical infrastructures. We followed Goodman's [31] guide for conducting a discourse analysis for psychologists - also drawing from Carla Willig's [87] and Locke and Budds [41] similar approaches. Our analysis first involved a preliminary reading with the primary researcher identifying the genre of the piece as well as the action orientation of language and which discourses are constructed and situated in the text. Next the primary researcher identified how language and discursive techniques such as positioning and framing was used in these discourses (Appendix A.1). At this stage, two further authors expanded and commented on these discourses suggesting new codes or offering alternative interpretations to existing codes. The primary researcher then derived resulting discourses relevant to the research questions iteratively along with feedback from the research team (Appendices A.2). We include further examples of our qualitative coding process in the supplementary materials.

### 3.3 Positionality

This research aims to avoid adding to the outside commentary and dissecting of celebrities' private lives. None of the research team make any claims of coming from an insider perspective of celebrity women. From our perspective as deepfake researchers, it is apparent that NSII is a problem that permeates every element of the technology. Every form of deepfake software we as researchers have interacted with has in some way ignored the consent of the target. Even the use of the term deepfakes across the literature immortalizes the Reddit user and community which popularized this content and perhaps should be re-considered as the dominant terminology.

## 4 ANALYSIS

Our analysis highlights the various ways in which victims frame the harms of deepfakes and how the infrastructures of social media and the wider internet enable abusers while challenging users' attempts to protect themselves/others and remove harmful content. First, we present findings relating to the three main factors from Baumer's [7] concept of Usees (lack of consent, lack of awareness and the experience of being targeted) and how these factors are constructed discursively with the experience of deepfake abuse. We then focus specifically on infrastructural and social factors which affect the experiences of deepfake victims.

### 4.1 How celebrity victims of deepfake abuse construct being *Usees* of deepfake technology

Our first research question focuses on how celebrities construct deepfake abuse as targeting them without their consent or awareness. Our analysis highlights how this experience is framed as a non-consensual manipulation of identity and how deepfakes are constructed as a form of non-consensual sex work. Repertoires of gendered violence are used to construct the targeted abuse of Usees and how they construct the distress of seeing this abuse imagery online as part of patterns of online harassment and shaming.

*4.1.1 Experiencing non-consent (deepfakes as manipulation of identity and as non-consensual sex work)*

Celebrity victims framed deepfake abuse as theft or non-consensual manipulation of their identity and constructed deepfake identities as non-consensual versions of themselves. They use possessive pronouns to highlight how despite the deepfake not being real, they own the faces in the content. "*Someone put my face onto porn and sold it*" (C). People constructed deepfake content as not only resembling them but as in some ways being them. "*I um was just shocked because this is my face, belongs to me*" (G). The action-orientation of deepfake abuse was framed in terms of stolen/manipulated identity, where deepfakes allow anonymous abusers to control their targets' identity and making it produce non-consensual sexual imagery. "*my face was stolen so men could make me into a sexual object to use for themselves*" (A). This is consistent with the role of non-consensual mimicry inherent in deepfake technology [3, 79]. This quote highlights one of many instances in the corpus where deepfake abuse was framed as a form of non-consensual online sex work. "*Porn of myself I never created has been viewed, distributed and sold. It has been used to threaten, humiliate and attempt to grab power from myself*" (E). E also draws from interpretative repertoires around gendered violence in how they construct deepfakes as non-consensual online sex work. In comparison to consensual online sex work, deepfake abuse is carried out without consent and distributed without permission (drawing parallels to having leaked videos and pirated content released).

Many statements used interpretative repertoires of stigma and reputational damage to highlight the inter-personal harms of non-consensual sex work online.

> "*Imagine that you found a video of you on the Internet where you were stamping on a kitten to death and everyone thought you were a complete asshole and people hated you and your boss was considering firing you and your family don't want anything to do with you and your friends are starting to leave and you don't know if your life is ever going to be the*

*same again and you don't feel safe walking down the street because now loads of people hate you because of something you never did. That's how it feels to be suddenly put in sex work when you never consented*" (C).

This example constructs the various different ways the reputational damage of deepfakes may affect the life of their targets. Multiple statements highlight how the fakeness of deepfake technology may not negate the harms of the technology to their reputation and personal relationships. "*My 60 year old dad if he were to see that video, I would never be able to convince him I didn't do it*" (QA). The reputational harms of deepfakes are fueled by repertoires of the believability of deepfake videos where false videos are framed as not being harmful.

*4.1.2 The experience of being targeted: repertoires of gendered violence*

The positioning of deepfake victims as non-consensual sex workers draws from repertoires around stigma and gendered violence. One statement situates the reputational damage of deepfakes as part of repertoires of sexual entitlement: the idea that expressing sexual autonomy is used as justification for non-consensual actions.

"*The minute that I seem sexually available in any way, shape or form, I'm fair game to humiliate physically assault, rape and attack. And this happens to porn stars. There are many porn stars who get murdered on their doorstep. Because a fan of their work will think that they're easy and that they will just sleep with anyone. So, there's this idea that the minute you are consensually sexual, everyone can help themselves to you and I've experienced this and I've seen it, and I've seen it happen to other people*" (C).

This quote constructs the harms of gendered violence against sex workers as motivated by interpretative repertoires of sexual entitlement and further constructs the potential dangers of non-consensual sex work. Other celebrities further situated this sexual entitlement in blaming narratives in cases where targets were harassed if they were seen as having "sexualized themselves" online.

"*[Famous streamer] was actually one of the people who was included in the streamers who were deepfaked. There were people coming into her chat saying "look at this photo or that photo of you that you posted". Trying to justify sexualizing her as if she brought it on herself*" (C).

This quote emphasizes how blaming narratives around deepfakes evoke stigmatizing attitudes towards sex, particularly those which tie into the values of sexual entitlement underlying "slut-shaming" tropes [66], which position women as responsible for their abuse.

Sexual entitlement was only one of many interpretative repertoires relating to gendered violence and targeted abuse online. Their statements situated deepfakes within existing forms of abuse and harassment specifically targeting successful women (and particularly women of color) "*It's really sick how y'all go out of the way to hurt me when you see me winning. Y'all going too far, Fake ass shit. Just know today was your last day playing with me and I mean it*" (I). Here, I positions the perpetrators of this content as specifically targeting her because of her success. Others framed this content within the Internet practice of targeting feminist celebrities/activists.

"*My first experience of this, my first knowledge of existing, was as a weapon to humiliate and degrade women. It was [Notable target of Gamergate] …and a popular streamer disliked her, so he basically mobilized his entire viewership against her. People started photoshopping her face onto pornography, and specifically sex acts that were considered degrading*" (C).

This discursive framing constructs deepfake abuse as targeted slander directed towards women who transgress or question societal norms, continuing a historical pattern of reputational damage which aims to undermine high profile women.

*4.1.3 The traumatic experiences of becoming aware of and watching deepfake abuse content*

Victims constructed social media as both providing a venue for targeted harassment and abuse, and for making targets aware of the abuse content. In particular, the malicious direct messaging of abuse content which forces streamers to see abusive imagery of themselves – often after the original case of abuse.

"*I just had my DMs open because I was looking for panel members for my award show. Yeah, so I opened it up for people to send me their stuff or whatever and I was like, huh, I'm just like sitting there, absorbing it and I click on my messages and it's just these videos*" (QA).

This example constructs the social harms of deepfake abuse, where outside of the direct creators of the content, social media facilitates the targeted bullying of victims by spreading and directly messaging abuse content. This highlights that while initially existing in communities which they do not interact with, the effects of deepfake abuse content begins to negatively affect the victims' ability to use social media wholesale. The public discourse around deepfake abuse incidents was where some victims became aware of their own imagery.

"*And when you see things like that, it's very hard not to sympathize and the whole situation led me to be a bit curious and be like, "wait am I on these websites?" so I went and googled it and unfortunately I had also been deepfaked by this person and you're about to learn how this is gonna fuck with my life forever*" (C).

This quote constructs the role of publicity and search engines as spreading awareness and serving as the first point of contact with traumatic content and abuse.

For those in the corpus who relayed first hand experiences of watching the deepfake abuse imagery, the content was framed in terms of various forms of sexual trauma and abuse. QA describes the psychological reaction to it as similar to traumatic experience of sexual abuse. "*Well, every woman that has been a part of this has kind of mentioned the same feeling. Like it reminds me of, like I was molested as a child. and it reminds me of that feeling like* (QA)". The traumatic experience of deepfakes similarly draws from interpretative repertoires around sexual assault:

"*In 2018, I was inebriated at a party and I was used for a man's sexual gratification without my consent. Today. I have been used by hundreds of men for sexual gratification without my consent. The world calls my 2018 experience rape. The world is debating over the validity of my experience today. This situation makes me feel disgusting, vulnerable, nauseous, and violated -and all of these feelings are far too familiar to me*" (A).

She frames the incident as directly comparable to the feelings of her traumatic experience of rape. It should be noted how the perpetrators are constructed here; while traditional assault often only has one direct perpetrator, deepfake abuse imagery is directly perpetrated by many individuals. Furthermore, the harm is exacerbated by the erasure and debate over the validity of the experiences of its victims

Another distinct element of the experience of encountering deepfake abuse imagery to note is how deepfakes evoke the harm of seeing an image or video of abuse which you do not remember happening to you, especially in cases where the content is more realistic. "*I saw my face like plastered on all these porn images and I was like Jesus I don't remember that, I don't remember that.*" "*Now, I can't be sure there's been some hazy no no haha. I'm adventurous- not that adventurous*" (H). This example parallels deepfake abuse with assault under the influence of alcohol. Finding realistic illicit deepfake videos of yourself online was framed as comparable to other forms of abuse imagery, particularly those which the victim does not remember. "*As soon as you see them, your mind shifts like from saying "Ohh. That's not my body why should I care" to – "Was I just? ...Did I do that? Do people have this of me?*" (B). This description of reconstructing her response to the content echoes commonly reported responses to other forms of gendered abuse and its impact on memory [33]: questioning first if the incident really happened and if the video is real.

### 4.2 How do victims of deepfakes negotiate infrastructural and social obstacles against seeking recourse for deepfake abuse?

Our second research question focuses on how infrastructural and social obstacles are negotiated by the victims of deepfake abuse. Firstly, we show how internet platforms are constructed as voicing narratives which blame and silence victims. We then note how deepfake infrastructure and public awareness of deepfakes are framed as further disqualifying the experiences of the abused. We then discuss how users construct the infrastructure of removing deepfake content and how it forces them to be responsible for removing NSII online. Finally, we return to the concept of public awareness and how it creates an ideological dilemma for advocacy.

*4.2.1 Constructing the internet and the disqualification of deepfake victimhood*

One fundamental part of the discourse around deepfakes and the internet was how the victims of deepfakes position themselves both as part of and outside of the internet. When positioning themselves as members of the internet, their existence in these online spaces was framed as justifying abuse as norm within this space, for example, "*yeah "What do you expect? You're part of the Internet"*" (B). The internet here is constructed as a community space where abuse and harassment are assumed as part of membership. They also position themselves as distinct from "the internet" and use speech to construct an "internet voice" which consists of the narratives being used to undermine their experiences:

> "*But if you look at a whole bunch of these comments, you'll see that so many people out there don't understand what the problem with this is… "Is there really any difference between this and some random viewer sexualizing them in my mind? I hate to break it to you, but you don't own the visage of your face. Log off. You're not a victim. What's the problem? Why do you care? This just comes with the job. Like this is normal. This is just a sexual fantasy. It's not that deep. Why do you care?… And there's a distinct lack of empathy and a lot of people who are genuinely confused as to what's the problem with it. And I'd say consent*" (C).

This quote constructs the various ways in which the action orientation of internet narratives undermine/disqualify the experience and emotional response of its victims: where a deepfake is framed as not real by comparing it to an act of imagination; victims frame themselves as having no ownership over their own image; their emotional reaction to the content is questioned; and the experience is framed as a byproduct of working as a streamer on the internet.

Victims often responded to the disqualification of deepfakes as being "not real" by situating the denial of deepfake victimhood within larger dialogues around consent. For example, G states: "*You know, we're having this gigantic conversation about consent, and I don't consent.*" She further rejects the position that labelling deepfakes as fake curtails harms associated with being deepfaked, "*so that's why it's not OK even if it's labelled as this is not actually her*". Victims further reject the disqualification of their experiences of victimhood through emphasizing how deepfake technology has impacted them psychologically. For example, B states: "*I think the really hard thing about it is like the reason I went live crying was because I get that so often, people being like, why does it affect them? I'm like, let me fucking show you*" (B). To challenge the framing of deepfakes as being a consequence of fame, celebrities construct the impact that deepfakes will have on society in general, and particularly on other vulnerable groups. F for example, states:

> "*Vulnerable people like women, children and seniors must take extra care to protect their identities and personal content. That will never change no matter how strict Google makes their policies. The Internet is just another place where sex sells, and vulnerable people are preyed upon*" (F).

This example constructs the internet as one of many unsafe places which facilitate the abuse of vulnerable people.

Despite the dominant positioning of deepfake communities who develop, support and spread deepfake content, celebrities also construct support from their online followers and other deepfake victims:

> "*I am relieved to finally be talking about this with some support. I was reached out to by 'myimagemychoice'. It's a group of very hardworking activists who are openly pushing for change and becoming a resource for the countless people who have been targeted and abused by others online*" (E).

This quote constructs a counter public of activists and supporters who are positioned as a protective online community space in contrast to deepfake abuse. This community support was especially vital in situations where the veracity of the deepfake video wasn't obvious and needed to be debunked. "*…so I appreciate everyone in the comments that were saying that, you know, it was fake and um supporting me. You guys are amazing*" (D). Some victims who were able to identify their abuser would reach out to their other victims to help other victims achieve justice: "*…she reached out to me and let me know. And she shared his identity with me. I now know his name, where he lives and where he works. And I am so ready to finally be able to pursue whatever action I can*" (E). Others organized group chats as collective spaces which they justified as emotionally supporting fellow victims. "*I want to fix stuff so I jump into action even though I'm crying, I get all these girls that were as many as I could. And I like dming them asking if they're OK*" (B). This quote highlights how in many cases the organization of communities of victims falls to victims themselves. It also constructs direct messages as a place for positive private interaction with other victims, in contrast to their use by abusers.

*4.2.2 Financial factors and lack of responsibility in deepfake recourses*
The celebrities in our corpus highlighted how the hosting of deepfakes contribute to the length and extent of the harms of NSII:

> "*You know what, the weirdest thing is you can't get them down like they're just so, anyone's allowed put them up and they are starting to look more realistic. Most recent ones are pretty realistic and I- I can kind of laugh about it, but at the same time I'm like, I hope that they come down at some stage. I don't want my kids to see them*" (H).

This quote highlights the role that internet platforms (specifically pornographic websites that host NSII) play in hosting and spreading defamatory content long after the original incident. Websites are constructed as both hosting this content and actively encouraging it by presenting users with advertisements of deepfake content. "*I've had deepfake porn pop up before and you see it and you're like, Oh my God. And you exit out and that's it. And you move on*" (B). This quote highlights how exposure to non-consensual deepfake content through pop-up ads framed as a common part of browsing the internet which users attempt to avoid – which has a disproportionate impact when victims of deepfake abuse are being exposed to advertisements for deepfake intimate imagery.

Victims of deepfakes highlighted the extent to which targets are forced to take responsibility for policing their own identity online and the various financial costs of this.

> "*Obviously, if a person has more resources, they may employ various forces to build a bigger wall around their digital identity. But nothing can stop someone from cutting and pasting my image or anyone else's onto a different body and making it look as eerily realistic as desired.*" (F).

This quote highlights the challenges of protecting a person's online identity from manipulation, and how this is furthered by the limited legal recourses available to victims. One person in the corpus focused on the hypocrisy inherent of platforms which host deepfake content produced by anonymous users (such as Reddit and twitter) do not face accountability for the anonymity which they provide and facilitate.

> "*There's a I believe it's a law called like number 203 or something like that and it protects all websites for what gets uploaded to their website. They're not responsible, the users are responsible, and you can't track the users because the Internet is anonymous. So yeah, hit a wall*" (B).

This quote highlights the dilemma for celebrity targets in the US, where the identifiable hosts of this content are immune from legal action, while they provide anonymity and a platform for abusers (as per Section 230 of US law with some exceptions including content which breaks US federal law or copyright law [15, 29]).

Just as various anonymous individuals/groups online profit from producing and advertising deepfake content, victims also framed deepfake recourses as a profit-motivated industry, where victims have to pay companies to clear harmful content which they struggle to remove themselves. For example, B states: "*I used to pay a company that cleans - websites for photos of you and- I said to him* [the perpetrator], *I said 'you are paying for every single girl to get clean'... I genuinely don't care because he caused the harm he has to fix it*" (B). The harms of deepfakes evoke repertoires of controlling one's own reputation online and the industry of removing content which often requires victims to pay for it. This example is interesting because the target puts the responsibility for this back onto a perpetrator, which they elaborate on in a subsequent interview. B would later state:

> "*The streamer who leaked the website has done a lot of work on the back end……he's really shown that he's trying really hard." He's partnered with a different AI company, actually, as ironic as that is, an AI company that combats, taking these*

*deep fakes down by using AI to find them and to delete them. So, you know, sometimes your worst enemy is your biggest help*" (B).

This example is unique in positioning of AI companies and deepfake perpetrators as both instigators of deepfake abuse and actively taking a role in removing the content, as part of their attempts to publicly apologize for their contribution to deepfake abuse.

*4.2.3 Everything you can do makes it worse: publicity, advocacy and reputational damage*

Engagement-based algorithms and publicity are framed as exacerbating reputational harms and encouraging silence; for example, search engine results privileging news coverage of deepfake abuse over their own content.

> "*and the reason I did a lot of the interviews is the same reason that a lot of girls didn't. A lot of girls don't want that to be the first thing you Google. Uhhh, you know, '[B] talks about deep fakes'. It does suck after years and years and years of building a career, that that's what comes up. And so, a lot of people opted out of doing interviews and in fact asked me or asked ahh [C], she's another one who's been vocal about it, to avoid that SEO attached to them, which is really sad. But I think it was important to talk about*" (B).

Through the framing of search engine results as something which is attached to victims, B specifically notes the effect of Search Engine Optimization (SEO) in how the victims of deepfake technology find their reputation and public perception affected. "*Fuck the fucking internet. Fuck the constant exploitation and objectification of women it's exhausting. It's exhausting. Fuck* [Male streamer] *for showing it to thousands of people. Fuck the people dming me pictures of myself from that from that website. Fuck you all*" (B). This extract highlights the anthropomorphic construction of the internet as a perpetrator, other internet celebrities as platforming and raising awareness of abuse sites and the use of direct messages of the content from other internet users as a distinct harm.

Publicity and coverage of deepfake abuse were also framed as privileging the experiences of abusers or outsiders over the experiences of the victims. "*The debate over our experience as women in this is, not shockingly, amongst men. None of you should care or listen to what any male steamer's "take" is on how we feel...This is not your debate. Stop acting like it is*" (A). Public apologies by creators and consumers of these materials after being caught with deepfake abuse imagery were framed as especially harmful as they further publicize the incident without any benefit to the target. "*It's just like so. It's so angry, anger inducing cause like the first thing he did was go live and apologize to his stream. He didn't apologize to any of the women, he apologized to his fucking stream*" (B). Deepfake publicity is seen as a way for the consumers of deepfake, not the targets, to mitigate the damage to their reputation. D encourages other internet users to engage with reporting mechanisms instead of further publicizing abuse content. "*I would really appreciate if anyone else ever see stuff like that to just report it instead of making videos about it.*" (D). This quote frames the difference between actions that reduce visibility of the content and those which publicize it.

*4.2.4 The ideological dilemma between the good of advocacy and the harms of publicity*

Often advocacy was seen as challenging the silencing of other victims. "*I know for a fact that so many others are suffering in silence. People are afraid to speak up, to talk about it, and I am so sorry that this happened to you too. I hope that I can encourage you that you are not alone*" (E). The function of this interview itself is advocacy, and it highlights various ways for victims and others to help enact legal change and to facilitate the advocacy of others. Because the organization and encouragement of advocacy is framed as a challenge to silencing narratives, many victims felt an obligation to speak up for other victims who were self-silencing. "*It's a tough issue and I had a sneaking suspicion maybe other people wouldn't want to talk about it and I feel a responsibility. I hope that we can continue conversations about this and see who it's negatively affecting. And help to change that*" (G). Many advocates continue to justify speaking out despite the harassment and publicity and see being an advocate as a responsibility.

> "*I'm willing to talk about this. And other girls aren't. And so I feel like I do need to be the one that's vocal because I'm willing to be that vocal one. So I'm trying to do. It's not only for myself, it's for a lot of other women. And that's I don't, I don't like, want to be vocal, but it feels like my only option. And so that's why I'm doing it*" (B).

This highlights a fundamental ideological dilemma where targets must reluctantly carry out advocacy. Public victims of deepfake abuse imagery are positioned both as victims and as expected to be advocates/activists.

As the first public victims of deepfake technology, they are obligated to take up a role of improving recourses for deepfake abuse. This dilemma is highlighted in the disparity between B's initial reactions to the content "*And the person that made that website. I'm going to fucking sue you, I promise you, with every part of my soul, I'm going to fucking sue you*" — In contrast with later discussions— "*it's just like, it's so frustrating to have to be like, OK, so now I want to sue the guy. So I've talked to three different lawyers trying to figure out how to sue the guy. They're telling me not worth your time. So now the only thing that I can do is try to create legislation. So now I have to meet with my my state legislator to try to get a bill passed of all. Blah all this stuff*" (B). This demonstrates the inherent power dynamics at play and the protection of website creators within current legislation. Because these celebrities position themselves as the first people to be targeted by this technology, taking an active role in helping develop new laws and adapt existing ones is framed as an obligation. However, speaking out and advocating on this issue opens them up to more reputational damage and harassment.

"*…there's all these media outlets that reached out to all of the girls involved saying, hey, do you want to do an interview? A lot of the girls are like, I don't want to do an interview because I don't want to further the narrative that I'm a part of this. That makes sense, but also without us doing interviews without more political outcry, then you don't get bills, you don't get legislation passed because politicians don't have the motivation, right? So so it's like I'm sitting here and I'm looking at this double-edged sword…*" (B).

This also highlights how the public coverage contributes to abuse and silencing which is sadly further justified in a later interview. "*But that blew up in my face too, so I don't know. I'm getting to the point, I think where I just um I just want to be more quiet, is what I'm learning. As sad as that is*" (B). There are two sides to deepfake abuse, the production and dissemination of the content and how the silencing and harassment of public-speaking victims discourages advocacy and justice.

## 5 DISCUSSION

This study analyses how the targets of deepfakes construct their experiences of lack of consent, lack of awareness and being targeted directly by technology. These constructions align with the characteristics of "Usees" as defined by Baumer [7]. Deepfakes are framed by celebrities as undermining consent in various ways, with targets positioned as non-consenting sex workers. Our analysis also highlights how the victims of deepfake abuse position themselves as targeted by abuse operating without their awareness, and they frame becoming aware of this abuse and the further harassment as part of broader repertoires of gendered abuse online. Celebrities construct their abuse as a distinct harm, involving being targeted and silenced by both by the creators and members of deepfake websites and a larger population of anonymous internet users who disseminate existing NSII, harass their victims, and subject them to stigma.

Our research highlights how deepfake abuse is enabled through infrastructure - emblemized by discursive constructions of cultures of harassment/silencing and limit recourses. Social media, the users of streaming websites and misogynistic groups that operate on the internet actively facilitate the creation and distribution of abuse content, as well as exacerbating the harms. The infrastructure of deepfakes is used to challenge victims across every stage of their abuse. Targets of this emerging technology must navigate their position as a victim with the obligation to activism dealing with further abuse. The platforms are constructed as not responsible for user-generated content while giving anonymity to, or even shielding, the creators of said content. Deepfake content removal is an industry; constructing other stakeholders who profit from deepfakes. This work has implications for our understanding of deepfake NSII and potential future work in this area.

### 5.1 Implications for Design, Policy and practice

Our study highlights how the burden of responsibility to address deepfake NSII is currently placed on victims of NSII abuse through analyzing the language they use in public statements. Taking an alternative, infrastructural, approach to understanding responsibility suggests that those with greatest responsibility for NSII, are those with infrastructural power and privilege [67]. In this instance, that is the deepfake content makers, the platforms that host and enable them, and the consumers of deepfake content. Understanding the lived experience of those who are targeted by deepfake technology is important, but it is not up to victims *alone* to make this system just. Based on our analysis, and to contribute to the critical discussion regarding responsibility towards deepfake content creation and dissemination, we propose several different stakeholders (see Table 2) as holding responsibilities for deepfake abuse and highlight potential ways HCI can challenge existing infrastructural harms.

Table 2: Various stakeholders identified as holding responsibility for deepfake abuse

| Stakeholders | Responsibility for harm | Potential for change |
|---|---|---|
| Social media platforms | Online celebrities rely on fan interactions in communities- but direct messaging is a venue for bad actors (who may not have the skill to create deepfakes themselves) to engage in targeted abuse. | Various services are now offering default blurring of videos with nudity (such as Google's safesearch and Apple's Sensitive content warning). This practice should be continued along with improved reporting and moderation policies and legal recourses. As well as interventions focusing distinctly on the targeted dissemination of deepfakes. |
| Search engines | Deepfake NSII is currently platformed on search engines. It is difficult to remove this content. | Deepfake abuse imagery websites should not come up in indexed search results, otherwise publicity on the topic will direct people towards abuse websites. Simply blurring the thumbnail is not sufficient; it helps people who do not want to view deepfakes but continues enabling those who do. |
| | Search engine optimization and engagement-based algorithms facilitate gendered violence and encourage the reputational damage of deepfake publicity by prioritizing content engagement. | Further research should be carried out on the impact of technology on reputation and how reputational damage encouraged by social media deepens the harms of gender-based violence. |
| | Generative AI often creates Usees of the technology because it is trained on and creates non-consensual data | Generative AI repeatedly creates Usees, undermining the values of consent and identity. This need to be a consideration in the ethical creation, application and regulation of generative AI |
| Online and traditional media outlets | For celebrities, the publicity around deepfake abuse is often the moment where they first realize, they have been targeted. | Trauma informed practice is vital in the public coverage of deepfakes and how news coverage is presented to people online in search engines and social media. |
| | Victims may feel forced to speak out about this topic or may not speak at all due to the cumulative effect of publicity and reputational damage. | There needs to be an increased focus on ethical journalism for victims of deepfakes to speak out anonymously and to not have their name publicly attached to abuse if they choose. Along with this, it is vitally important to protect those who do want to speak out from further targeted online abuse. |

Current deepfake recourses encourage targets to remove their presence from social media, as opposed to removing the anonymous perpetrators, whose identities are protected [43]. This paper does not seek to tackle the various competing epistemologies around issues of anonymity, freedom and privacy online (which have been well-discussed previously [2]. However, designers should at least consider how we can offer support for targets of deepfake technology to maintain and protect their own public identity in the face of harassment and reputational damage. People who are encouraged to engage with social media as part of their professsion already need to balance visibility online with the various risks of having their identity mis-represented [18]. Existing literature has highlighted the personal, social and platform vulnerabilities of being public online and the failure of social media companies to adequately protect online users [19]. Social media sites and search engines need to augment their current reporting mechanisms and allocate appropriate resources towards the removal and prevention of abusive content. There are numerous ways this could be facilitated: treating AI generated pornography as pirated content (as the systems behind enforcement and detection of copyright infractions online are much more efficient) [64] or other forms of illegal pornography [51] and hindering its discoverability or transmissibility online; or reconsidering how the names of victims become attached to deepfake content and publicity in search results. Outside of the implications for deepfake recourses, our work has broader implications for the design of technology that creates Usees, and the how those Usee-centered technologies erode consent. The informed consent of the targets of a technology is fundamental to whether one is designing for Usees or users. The value of consent needs to be implicit in design of deepfake technology, not an afterthought, and not something to be retroactively gathered. In the same way we consider who technology will be used by, we must also ask ourselves who technology may be used on. Otherwise, this same discussion will be had again, and again as new technologies create more non-consenting targets.

To progress the emerging infrastructural understanding of deepfake harms [67], our study highlights the industry that supports and depends on deepfake abuse. Deepfake marketplaces produce deepfakes for profit; pornographic websites profit from the distribution of the imagery; internet search providers profit off search traffic; and websites with no direct relationship to deepfake content may host deepfake-related advertisements and thus funnel users towards these harmful sites. Publicity around deepfakes often furthers the profits of all parties involved. The experience recounted by the celebrities in the corpus highlights that recourses against deepfake abuse are repeatedly infrastructurally challenged by the industry of deepfakes. The removal of abuse content for example is a profitable industry in its own right, preying off users' inability to remove content of themselves. Victims of deepfake abuse are being encouraged under the current system to gather abuse evidence themselves, with no guarantee of any protection from future abuse, and this process of removal must be manually repeated for every major search engine and platform [65]. While it's useful that for-profit reputation management can offer (sufficiently wealthy) celebrities some element of control over their public identity, these services must be repeatedly used (and paid for) while NSII abuse continues unchecked. Infrastructural justice needs to prioritize the equitable treatment of deepfake targets and access to recourse over profit.

### 5.2 Implications for HCI: Reconsidering deepfakes through the lens of trauma informed computing

Existing work in HCI on trauma informed computing [14] has highlighted personal trauma as a lens for understanding and designing interactions with technology. Our work details the distressing first person accounts of deepfake NSII across every stage of their abuse and attempts to seek recourse. Drawing on the framework of trauma-informed computing put forward by Chen and colleagues [14], we draw from the four Rs of trauma informed care [35] to identify areas in which trauma informed computing may challenge deepfake abuse: realizing trauma exists, recognizing trauma, responding to traumatic content and resisting re-traumatization (see Table 3). There are cultural differences in the recourses for deepfake abuse available to victims and the recourses available have changed over time. In many ways our dataset reflected celebrities' response to increasing regulation and support services becoming available. Still, existing work has highlighted how victims from different countries may not be aware of what support and recourses are available in their country [81] – this is something which also needs to be taken into consideration when designing support services for deepfake NSII.

Table 3. Implications of our work for Trauma Informed Computing targeting deepfake NSII

| Principle of trauma informed computing | What this means for deepfake NSII |
|---|---|
| **Realizing that trauma exists** | It is vital for HCI researchers to understand that deepfake videos can be traumatic. Both to design trauma informed supports services for victims of NSII and to challenge silencing narratives online. |
| **Recognizing signs of trauma** | We expand on Baumer's concept of unaware Usees by recognizing how the unique distress when one becomes aware they have been targeted by deepfakes. How Usees become aware (social media, news publicity or harassing messages) is a significant point of first contact between Usees and abuse materials |
| **Responding to the content by applying the principles of trauma informed care** | Challenging the role of social media in facilitating NSII:<br>    Social media fosters abuse by connecting abusers to each other and abuse technologies and serving as a venue to directly harass and defame individuals. There needs to be work on facilitating infrastructure which supports victims of NSII. The positive infrastructure identified include support groups established by victim-survivors and the mobilization of internet users in reporting the content and hindering visibility.<br><br>Mitigating the trauma of becoming aware you've been deepfaked:<br>    There is some evidence that prompt-based interventions on search engines can be successful in directing individuals towards help and support [5]. For deepfake abuse, existing social recourses and content removal options should be prompted at users whose search terms include the word deepfake along with any persons' name. |
| **Resisting re-traumatization** | The distress of deepfakes is compounded by three factors which may contribute to re-traumatization:<br>    1. Dealing with the publicity of deepfakes and public debate |

2. Increased harassment either by shaming or silencing or sending deepfake videos directly to their targets through DMs
3. Sharing of old deepfakes after original incidents or creation of new deepfakes

Existing research on trauma informed design highlights how social media has the potential to exacerbate various forms of trauma [69]. Further research is needed which considers NSII and other forms of technology facilitated gendered violence, e.g., developing communities online for victims, improving moderation online and adapting existence support communities for gendered violence to support harms from synthetic imagery. We also call for an increased focus on how the design of technology and social media has the potential to fuel trauma and hinder victims' ability to respond or find recourse to abuse. The most significant part of this approach is that HCI researchers need to consider the role of generative AI in image-based sexual abuse, how anonymous networked misogyny can organize around furthering abuse and sharing abuse materials and how direct messaging can serve as a venue for harassment using synthetic abuse imagery. Researchers need to consider the interactions between celebrities and their fans on social media and how the promises of connection to fan communities is also allowing abusers a chance to directly harass celebrities.

There are further implications of trauma-informed approaches to how we carry out research with deepfake abuse survivors in a trauma informed way. Razi and colleagues highlight the importance of trauma informed research in carrying out interviews with teenagers on their experience with social media and the discomfort brought up by discussing harmful experiences with technology [62]. Some of their recommendations must be considered in light of our own findings when carrying out research which seeks to engage with victims of deepfakes and media editing technology. For example, when *anticipating knowledge gaps* and *anticipating emotional vulnerabilities* of research participants, we must account for the possibility of victims becoming aware that deepfakes or other similar edited media may exist of them which our dataset highlights as a particularly distressing experience.

## 5.3 Implications for implementing policy and interventions

Most classifications of deepfake perpetration classify the abuse into three stages involving creation, dissemination and removal [21]. The celebrities in our corpus construct a fourth phase: debate. The final stage of deepfake abuse involves varying social and reputational harms arising from the publicity of and social aftermath of NSII. There are prescribed ethical standards for reporting on sexual assault such as including helplines and not revealing the identity of the target, though these standards are often ignored [4]. We recommend that publicity that raises awareness of deepfakes should also raise awareness of existing recourses and support. It should advocate for the victims as opposed to giving publicity to the creators of deepfakes and viewers of deepfakes. Like celebrities, activists are a group who are often targeted specifically by deepfake abuse and harassment [43]. When celebrities take on the dual role of being celebrities and also public activists against deepfakes they are doubly subject to further targeted harassment by misogynistic internet users/groups. It is important to recognize that the advocacy provided by deepfake victims is essential to challenging the blaming and silencing narratives around the abuse. But public advocacy needs to be a choice over an obligation, especially while advocacy alone can't challenge deep rooted obstacles towards justice [52].

A major contribution of this work is the need for intervention at every stage of deepfake abuse, targeting various stakeholders. Beyond challenging the creators of deepfake abuse imagery, work needs to directly confront the rationalization and motivations of those who participate in the production and dissemination of deepfake abuse. Existing HCI work has shown the role that technology can play in engaging with the perpetrators of gendered abuse [9]. The motivations not only for the production/consumption of deepfakes but the distribution of content and debate of the experiences of women should be challenged by future intervention work. Our research shows for the first time how online culture around the production and accessibility of deepfakes create discourses that falsely minimizes harms to justify current and future abuse (Table 4). The existence of these deepfake NSII myths needs to be considered as a variation of existing myths around various forms of gendered violence [23] which have the potential to lower rates of reporting [26]. The design of perpetrator interventions for NSII myths should draw from existing HCI research into automated and manual interventions for challenging rape myths [56, 75].

Table 4. Deepfake NSII myths from our corpus and how they challenged

| Deepfake NSII myth identified in our corpus | How do celebrities challenge these myths in our corpus |
|---|---|
| 1. Victim-survivors deserve deepfakes because they are public figures/women on the internet | Abuse imagery should not be part of someone's job |

| | | |
|---|---|---|
| 2. | The content isn't real, so it can't harm | The distress of seeing a deepfake is real and the social harms of the technology are also very real |
| 3. | If the content is labelled as false it isn't harmful | Being labelled as fake doesn't rectify the non-consensual and disturbing nature of the imagery |
| 4. | Victim-survivors are pretending to be affected by deepfake content | Deepfake victim-survivors are affected by the content in multiple different ways |
| 5. | Celebrities who are perceived as producing sexualized content should expect deepfakes | Deepfakes can be made of anybody, this argument is motivated by sexual entitlement |
| 6. | Deepfake abuse imagery is just a "joke" | Deepfake abuse imagery has serious personal and social consequences |
| 7. | There is no difference between making abuse imagery and having a sexual fantasy about someone else | Sexual fantasies aren't distributed on the internet, aren't being used to harass celebrities, aren't based off training sets designed using celebrity images |
| 8. | Deepfake victim-survivors don't have any ownership of their face/image/body | Just because someone's image may not be protected by the law doesn't mean their image is not theirs |

## 6 LIMITATIONS

The main limitation of this work is that the sample is limited to nine western celebrities. Our justification for focusing on this sample of victims was that they are in many ways, a "canary in the coal mine" in regard to deepfake abuse. Their experiences may inform how we approach deepfake abuse as a growing problem targeting more and more victims. Just as stolen sex tapes and non-consensual imagery of celebrities online emerged alongside the growing crisis of online IBSA targeted towards the average person. It also been remarked that online culture has also brought with it more comparable experiences of public repreentation between the average person and celebrity and that the IBSA myths faced by celebrities will eventually target the non-celebrity [44]. Deepfakes are becoming more accessible, the realism and quality offered by publicly available consumer technologies has grown rapidly [53] which may lead to an increase in perpetration. There have been an increasing number of deepfake NSII targetting members of the general population, even towards schoolchildren [6] . However, we also take this opportunity to re-affirm the necessity of supporting celebrity women as victims in their own right. While celebrity women in many ways have more social privilege than non-celebrities and reflect on this in their statements, their experience of reputational damage is particularly harmful as it challenges their livelihood. Nonetheless, it is also important to note that the infrastructures of gendered violence discourage many more marginalized groups from equal participation in online activism [52] and so the individuals included in our sample are likely not representative of all victims of NSII. With the increasing accessibility of deepfake technology, our work serves as a call for future research to take a more holistic view of deepfake targets, exploring how the intersection of publicity, class, gender, sexuality, occupation and race may radically affect experiences of deepfake abuse.

## 7 CONCLUSION

Victims of deepfake NSII are an example of Usees of technology as they are targeted by a technology without consent or awareness. Our research highlights how high-profile celebrity targets construct being targeted by deepfakes and their interactions with existing legal, social and technological hurdles to finding recourse. As HCI researchers, we should challenge technology which creates Usees. We encourage critical consideration of how the manipulation of identity is baked into existing generative AI models. This work challenges the current infrastructure of deepfakes: the failures of the limited resources available has led to deepfake removal becoming a for-profit industry. As victims of a novel technology, celebrities are often obligated into advocacy which increases the abuse and reputational damage. Future research should take a trauma informed lens, and we provide recommendations on how to report and publicize deepfake abuse while challenging myths about NSII and its victims.

## ACKNOWLEDGEMENTS
This research was conducted with the financial support of Taighde Éireann– Research Ireland under Grant number 13/RC/2094_2

# A  APPENDICES

## A.1  Analysis Example

Partial extract of one of the texts and its analysis – we would've preferred to reproduce the full texts for each interview in the corpus but reproducing entire texts would be outside the bounds of fair use and thus represent a copyright issue.

| Text | Genre | Action orientation, situation, construction | Discourse elements |
|---|---|---|---|
| … Even among my peers, so many of us have had these same exact experience. And it kills me to know that this is an inevitability that future generations will also be victim to. And even worse, because technology advances so quickly. Right now, the laws surrounding deep fake porn and other intimate image abuse are almost non-existent. It has felt hopeless and honestly unsafe to ever speak out. However, I am relieved to finally be talking about this with some support. I was reached out to by 'My image. My choice'. It's a group of very hardworking activists who are openly pushing for change and becoming a resource for the countless people who have been targeted and abused by others online. So I am finally looking into what I can do, especially because I received a very interesting message recently and that's the other part of my story that I wanted to address. A lot of people wrongfully think that deep fake porn is a joke, or that because I post my face publicly, I deserve this. But it can happen to anyone. I was contacted by someone who thinks she has actually discovered the identity of a person who has created deep fakes of her. This person hiding behind a username on one of the biggest deep fake sites was recognised because he made pornographic videos of at least seven people he knows, personally, from his time at school, so not people in the public eye, but classmates. But this same username also made a tonne of videos of me and a dozen of my friends. So she reached out to me and let me know. And she shared his identity with me. I now know his name, where he lives and where he works. And I am so ready to finally be able to pursue whatever action I can. To find some semblance of justice on behalf of all the people he's done this to. Unfortunately, he's obviously just one of many, many offenders who have exploited others in this way. But to me it does feel like a huge step closer in finally fighting back because I don't care if it's just one. I want to let him know that I know. This affects people's lives. It's fully ingrained in mine. It dictates how I act, where I feel safe and is something that I'm forced to come to terms with in order to just continue existing. | (G) Advertising<br><br>(G) Narrative | (A) Highlights group experience of deepfakes<br><br>(C) Community of targets which is rapidly expanding<br><br>(A) Justifying silence as safety<br><br>(C) Constructs internet people who are shaming<br>(S) responses situated on timeline before and after identifying the perpetrator, changes in self over time<br>(C) Construction of community of targets in solidarity, construction of the deepfake creator, non celebrity targets<br><br>(C) Anonymity and identity<br><br>(C) Constructs Deepfake recourse | (P) Positions the experiences as belonging to self and other (My and us), both a personal experience and a communal narrative<br><br>(ID) Balancing want to speak out with potential for more harm, and lack of actual consequences<br><br>(P) positioning self within a popular shaming narrative that public celebrities as being uniquely targeted because of lack of image rights, (F) lack of self framed as trade-off of celebrity, power disparity between public figure and anonymous users<br><br>(F) Anonymity as disguise, username as masks<br><br>(L) Not just public figures but individual known targets<br><br>(F) Loss of anonymity of perpetrator - evening out power disparity, action is something which is owed to other victims<br>(F) Deepfakes as having large amount of anonymous perpetrators<br>(F) Deepfake recourses as a stepped process (see[Co-auuthor name] comment on back to square one)<br>(F) Deepfakes affecting multiple different domains of life<br>(GR) (F) "It" is privileged over targets behaviour and safety, framed as inescapable/ a requirement of continuing ones life |

## A.2   Example of writing up the analysis

**See supplementary materials for a full list of codes in the dataset (S.1), early drafts of discourse and an example of the process of organising the codes by discourse (using colour) (S.2) and the organisation of relevant quotes by discourse (S.3)**

*A.2.1 Example paragraph from early write up*

**Community values/attitudes encourage silence/blaming**

**Public coverage facilitating harms and traumatisation Internet control over narrative -speaking to the internet**

In our dataset, victims position themselves and experience but also construct the public discourse on deepfake abuse imagery and frame publicity as drowning out the experiences of victims while privileging the experiences of outside commentators and perpetrators. One distinct part of the public coverage which emerges in relation to online celebrities such as streamers is how the reactions of the perpetrators and other male streamers are privileged over the experiences of the victims. "The debate over our experience as women in this is, not shockingly, amongst men. None of you should care or listen to what any male steamer's "take" is on how we feel"…… "This is not your debate. Stop acting like it is." This example shows the inconsistency where the coverage of this abuse overly focuses on commentary and reactions over lived experience. The focus on perpetrators also came up in regard to public coverage of deepfakes "I don't know. I think what's frustrating is because I had to do. It's just like so. It's so angry, anger inducing cause like the first thing he did was go live and apologise to his stream. He didn't apologise to any of the women, he apologised to his fucking stream. And it's like what are you doing?" Public apologies to the general public were framed as especially harmful as it furthers publicity about the incident without any benefit to the target. "I think I could forgive if he would have like called me on a Tuesday and he's like, hey, I looked at this three weeks ago. I'm so sorry. I think I could find a place and like, I could forgive him for that. But the problem is that the platform and the wildfire that it's spread is just like, it feels so unrepairable at this point". Two distinct harms are framed here, consumption of deepfakes and contributing to the spread of deepfake content. The spreading awareness of the platform and causing the stressful public coverage highlights the role of publicity in the harms of deepfakes. The news publicity around deepfake abuse incidents was also identified as a moment where some targets first became aware of their own imagery. "And when you see things like that, it's very hard not to sympathise and the whole situation led me to be a bit curious and be like, "wait am I on these websites" so I went and Googled it and unfortunately I had also been deep faked by this person and you're about to learn how this is gonna fuck with my life forever." The role of technology and search engines as first point of contact with trauma and abuse is important to note here as. Outside of the direct consumers and disseminators of deepfake imagery, the spreading of awareness of the imagery is also framed as contributing to the harm. "I'm. I'm so disgusted. I mean, I guess it's giving you clout. I guess. You got 2.6 million views, so It's a great way to go viral. I mean, I guess. I've never met you. I'm sorry if anything I've ever done has upset you, but you went out of your way to put hashtags about me being a pickme and leaks and stuff when it took two seconds to know that those are fake and really dehumanising to do to another woman." She expands on this further: "It's really hurtful to see another young woman be so dehumanising because it took two seconds to know that those are fake instead of spreading that everywhere and me getting disgusting umm DM's about it, which I really don't want to see. I already had to take a couple days off of um social media this week because of the way that people um sometimes message me and it's just really disgusting." In this example, another influencer is positioned as contributing to deepfake abuse by spreading awareness about "leaked" images, leading to the victim being harassed and having to take time off from the internet. The streamers in the dataset particularly highlight how the public awareness feeds into patterns of online harassment and having their experiences disqualified You get accused of things like every day and you get called names every day and you just get to a point where you say," OK, I don't. I don't. I can't care". / I forget that there's no humanity on the Internet. This quotation also highlights how the targets position themselves as part of an uncaring internet, aimed on challenging their experiences. We will further discuss the characterisation of the internet and how it is used to voice shaming/silencing narratives around deepfake abuse.